\documentclass[final,3p,twocolumn]{elsarticle}

\usepackage{graphicx}
\usepackage{color}

\usepackage{amssymb}

\journal{Astroparticle Physics}

\begin{document}

\begin{frontmatter}

\title{Radioactivity backgrounds in ZEPLIN--III}

\author[ICL]{H.M.~Ara\'{u}jo\corref{cor1}}
\cortext[cor1]{Corresponding author}
\ead{h.araujo@imperial.ac.uk}
\author[ITP]{D.Yu.~Akimov}
\author[EDI]{E.J.~Barnes}
\author[ITP]{V.A.~Belov}
\author[ICL]{A.~Bewick}
\author[ITP]{A.A.~Burenkov}
\author[LIP]{V.~Chepel}
\author[ICL]{A.~Currie}
\author[LIP]{L.~DeViveiros}
\author[RAL]{B.~Edwards}
\author[EDI]{C.~Ghag}
\author[EDI]{A.~Hollingsworth}
\author[ICL]{M.~Horn}
\author[RAL]{G.E.~Kalmus}
\author[ITP]{A.S.~Kobyakin}
\author[ITP]{A.G.~Kovalenko}
\author[ICL]{V.N.~Lebedenko}
\author[LIP,RAL]{A.~Lindote}
\author[LIP]{M.I.~Lopes}
\author[RAL]{R.~L\"{u}scher}
\author[RAL]{P.~Majewski}
\author[EDI]{A.St\,J.~Murphy}
\author[LIP,ICL]{F.~Neves}
\author[RAL]{S.M.~Paling}
\author[LIP]{J.~Pinto da Cunha}
\author[RAL]{R.~Preece}
\author[ICL]{J.J.~Quenby}
\author[EDI]{L.~Reichhart}
\author[EDI]{P.R.~Scovell}
\author[LIP]{C.~Silva}
\author[LIP]{V.N.~Solovov}
\author[RAL]{N.J.T.~Smith}
\author[RAL]{P.F.~Smith}
\author[ITP]{V.N.~Stekhanov}
\author[ICL]{T.J.~Sumner}
\author[ICL]{C.~Thorne}
\author[ICL]{R.J.~Walker}
\address[ICL]{High Energy Physics group, Blackett Laboratory, 
Imperial College London, UK}
\address[ITP]{Institute for Theoretical and Experimental Physics, 
Moscow, Russia}
\address[EDI]{School of Physics \& Astronomy, SUPA University of Edinburgh, UK}
\address[LIP]{LIP--Coimbra \& 
Department of Physics of the University of Coimbra, Portugal}
\address[RAL]{Particle Physics Department, 
STFC Rutherford Appleton Laboratory, Chilton, UK}


\begin{abstract}

We examine electron and nuclear recoil backgrounds from radioactivity
in the ZEPLIN-III dark matter experiment at Boulby. The rate of
low-energy electron recoils in the liquid xenon WIMP target is
0.75$\pm$0.05~events/kg/day/keV, which represents a 20-fold
improvement over the rate observed during the first science
run. Energy and spatial distributions agree with those predicted by
component-level Monte Carlo simulations propagating the effects of the
radiological contamination measured for materials employed in the
experiment. Neutron elastic scattering is predicted to yield
3.05$\pm$0.5 nuclear recoils with energy 5--50~keV per year, which
translates to an expectation of 0.4~events in a 1-year dataset in
anti-coincidence with the veto detector for realistic signal
acceptance. Less obvious background sources are discussed, especially
in the context of future experiments. These include contamination of
scintillation pulses with Cherenkov light from Compton electrons and
from $\beta$ activity internal to photomultipliers, which can increase
the size and lower the apparent time constant of the scintillation
response. Another challenge is posed by multiple-scatter $\gamma$-rays
with one or more vertices in regions that yield no ionisation. If the
discrimination power achieved in the first run can be replicated,
ZEPLIN-III should reach a sensitivity of
$\sim$1$\times\!10^{-8}$~pb$\cdot$year to the scalar WIMP-nucleon
elastic cross-section, as originally conceived.

\end{abstract}

\begin{keyword}
Liquid xenon detectors \sep dark matter searches \sep ZEPLIN-III \sep
radioactivity \sep background studies
\end{keyword}

\end{frontmatter}


\section{Introduction}
\label{intro}

\begin{figure}[ht]
  \begin{center}
    \includegraphics[width=7.8cm,clip=on]{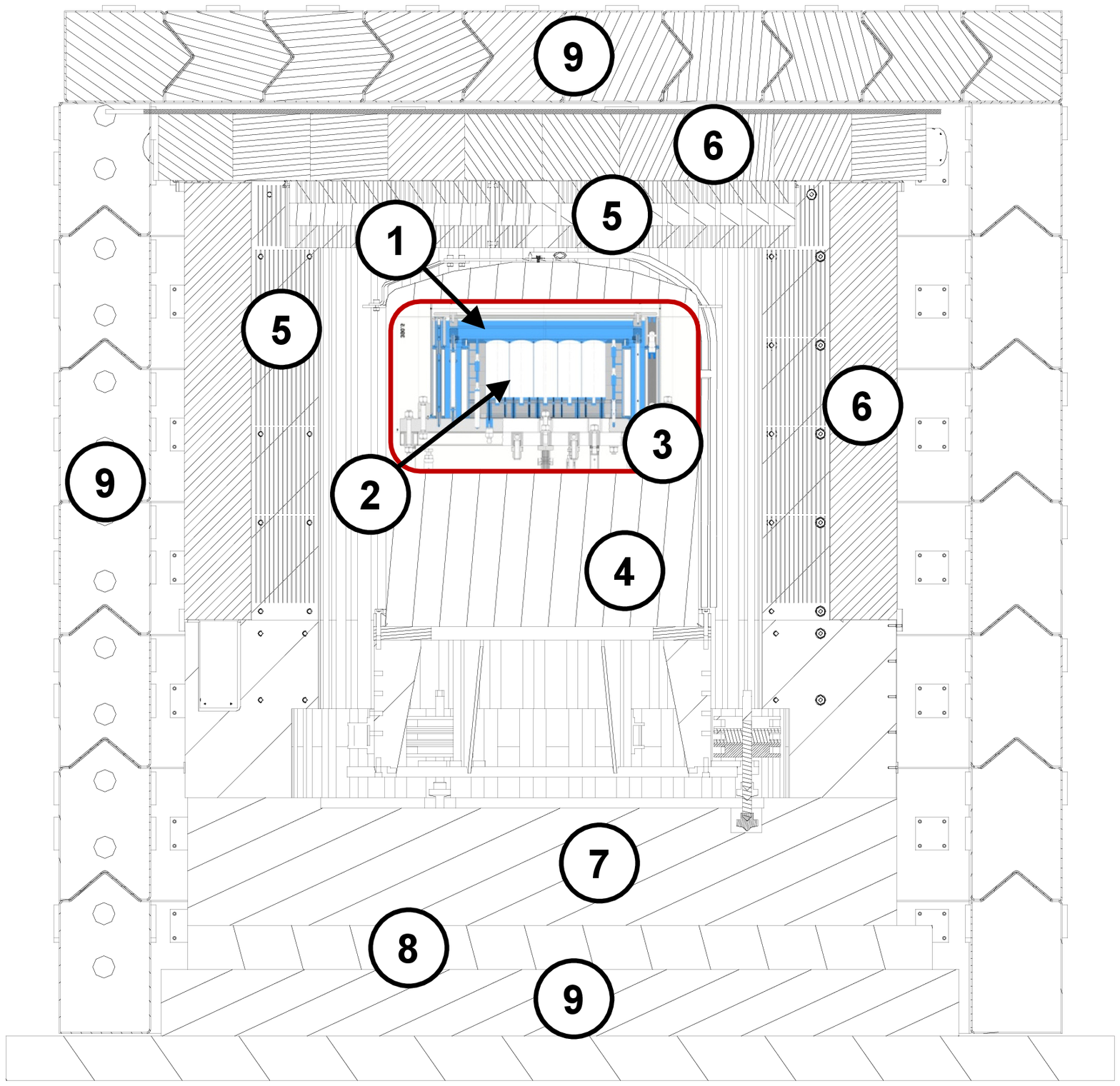}
  \end{center}
  \caption{\label{z3} Schematic diagram of the ZEPLIN-III
  experiment. 1: WIMP target (LXe shown in blue); 2: photomultiplier
  array; 3: xenon chamber; 4: vacuum cryostat; 5: Gd-loaded
  polypropylene; 6: veto scintillator modules; 7-bare polypropylene;
  8-copper flooring; 9-lead shielding.}
\end{figure}

The ZEPLIN-III direct dark matter search has been operating at the
Boulby underground laboratory (UK) under a rock overburden of 2,800~m
water equivalent. The WIMP target is a xenon emission detector
\cite{dolgoshein70} which discriminates between electron and nuclear
recoils by measuring the relative amount of prompt scintillation light
and ionisation charge extracted from particle interactions in the
liquid xenon (LXe) phase; the ionisation is transduced into a second
optical signal via electroluminescence in the thin vapour layer above
the liquid. Both the primary (S1) and secondary (S2) scintillation
pulses are detected by an array of 31 photomultipliers immersed in the
liquid. Key components of the experiment are shown in Figure~\ref{z3};
for details on the design and construction of the system we refer the
reader to Refs.~\cite{araujo06,akimov07,akimov10b,ghag11}.

After deployment underground in early 2007, an 83-day run produced
competitive limits for WIMP-nucleon scattering cross-sections
\cite{lebedenko09a,lebedenko09b,akimov10a}. In this first phase the
background of the experiment was dominated by the photomultiplier
array by a significant factor -- both in electron recoils from
$\gamma$-rays and in nuclear recoils from neutrons. A development
programme was thus undertaken with industry in order to develop new
phototubes targeting an order of magnitude reduction in
radioactivity. After upgrade of the photomultipliers and the addition
of a veto system around the main target \cite{akimov10b,ghag11},
the experiment returned to taking WIMP search data in mid 2010. The
first aim of this article is to describe the radioactivity backgrounds
which determine the sensitivity of the rare event search ahead of
forthcoming results from the second run.

We highlight also that many background studies focus solely on simple
event topologies quantifiable by Monte Carlo simulation, but often
overlook other important factors. Although the requirement of
negligible neutron background drives the design of most dark matter
experiments, in fact this threat materialises only rarely. In
two-phase xenon detectors, multiple scattering $\gamma$-rays can be
challenging when one of the vertices occurs in a region which yields
no ionisation; in this instance the prompt scintillation signals for
the multiple vertices are time-coincident and difficult to
distinguish, and the charge-to-light ratio (S2/S1) is low, more
typical of nuclear recoils. In another example, when estimating
photomultiplier backgrounds, many studies consider only high-energy
$\gamma$-rays as contributing significantly to electron recoil
background; we present evidence here that $\beta$-induced signals
generated internally (involving, for example, Cherenkov emission from
the phototube envelopes) should also be assessed carefully.

Besides their sensitivity in the scintillation and ionisation channels
(see,~e.g., Ref.~\cite{edwards08} for an emphatic demonstration of the
latter), two-phase xenon detectors hold great promise for next
generation dark matter searches due to the ability to self-shield
against external backgrounds. The accurate reconstruction of the
position of interactions in three dimensions allows a sacrificial LXe
volume to be defined which shields an inner `fiducial' mass from most
sources of radioactivity. ZEPLIN-II was the first WIMP experiment to
operate using this technology \cite{alner07}, soon followed by XENON10
at Gran Sasso \cite{angle08} and ZEPLIN-III at Boulby
\cite{lebedenko09a}. XENON100~\cite{aprile10} already benefits from
self-shielding very significantly, and even more so will the upcoming
LUX350 experiment at SUSEL~\cite{mckinsey10}. Two-phase xenon
experiments have been at the forefront of WIMP sensitivity and systems
with tonne-scale fiducial mass could probe most of the parameter space
favoured by constrained minimal supersymmetry (cMSSM) and by other
extensions to the standard model. With this work we wish to summarise
a decade of experience in designing, building and operating
ZEPLIN-III, and thus inform the design of next-generation WIMP
experiments.

This article is organised as follows. Ordinary, single-vertex electron
recoil backgrounds are treated in Section~\ref{gammas}, where
ZEPLIN-III data are confronted with an overall prediction built up
from component-level simulations. In Section~\ref{odd} we discuss
event topologies that may degrade the discrimination power of these
instruments, such as $\gamma$-ray multiple scatters and $\beta^-$
background internal to the photomultipliers. Having validated U/Th
contamination levels in key components, we present calculations of
neutron background in Section~\ref{neutrons}. Finally, we summarise
our findings in Section~\ref{conclusion}, with the design of future
systems in mind.

\section{Single-vertex electron recoil backgrounds}
\label{gammas}

Most low-energy electron recoils in the LXe result from interactions
of $\gamma$-rays and $\beta$-particles from natural radioactivity. We
begin by calculating absolute spectra of deposited energy from
dominant sources, and compare these with ZEPLIN-III data acquired at
Boulby in fully-shielded configuration. These signals can be
discriminated from nuclear recoils to a large degree; it is reasonable
to expect that a `leakage' (mis-identification probability) of order
1:10,000 may be achieved for single-site interactions such as those
from $^{85}$Kr $\beta$-particles within the LXe (the value eventually
achieved in the first run for all interactions was 1:7,800
\cite{horn11}). However, more complicated topologies must be
considered, and these may not be rejected so efficiently. We analyse
two such types in Section~\ref{odd}.

The $\gamma$-ray background is dominated by primordial U, Th and
$^{40}$K. We assume natural uranium (0.956~$^{238}$U, 0.044~$^{235}$U)
and that all chains are in secular equilibrium, except where indicated
otherwise. To determine contamination levels, a large number of
materials were radio-assayed over the lifetime of the project. All
HPGe $\gamma$-ray spectroscopy was conducted at Boulby with an ORTEC
GEM detector with a 2~kg crystal, reaching down to $\sim$0.5~ppb
sensitivity for U and Th. Energy-dependent counting efficiencies were
calculated from Monte Carlo simulations for each test geometry. These
measurements were complemented by commercial mass spectrometry
(ICP-MS) for several materials. We have also benefited from the
extensive UKDMC database \cite{ukdmc}.

The 20 most intense $\gamma$-rays in each chain and the 1,461~keV line
from $^{40}$K are propagated using the GEANT4 Monte Carlo
toolkit~\cite{agostinelli03} to interactions in the LXe target. For
internal backgrounds the ZEPLIN-III simulation model was
used~\cite{araujo06}; this was complemented by importing a detailed
CAD solid model for calculations involving the veto, shielding and the
laboratory rock.

\subsection{ZEPLIN-III measurements}

\begin{figure}
  \begin{center}
    \includegraphics[width=7.8cm]{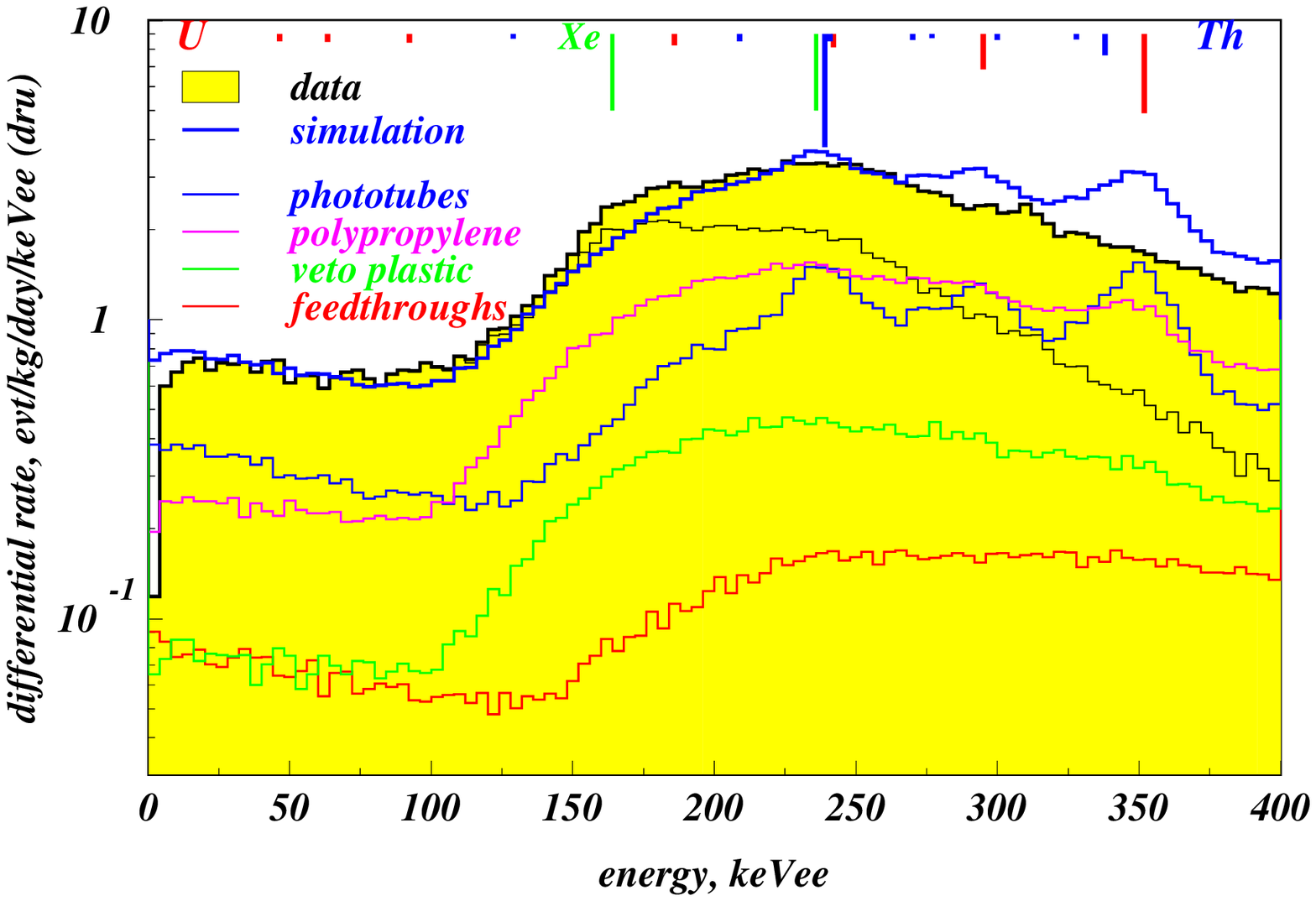}\\
    \includegraphics[width=7.8cm]{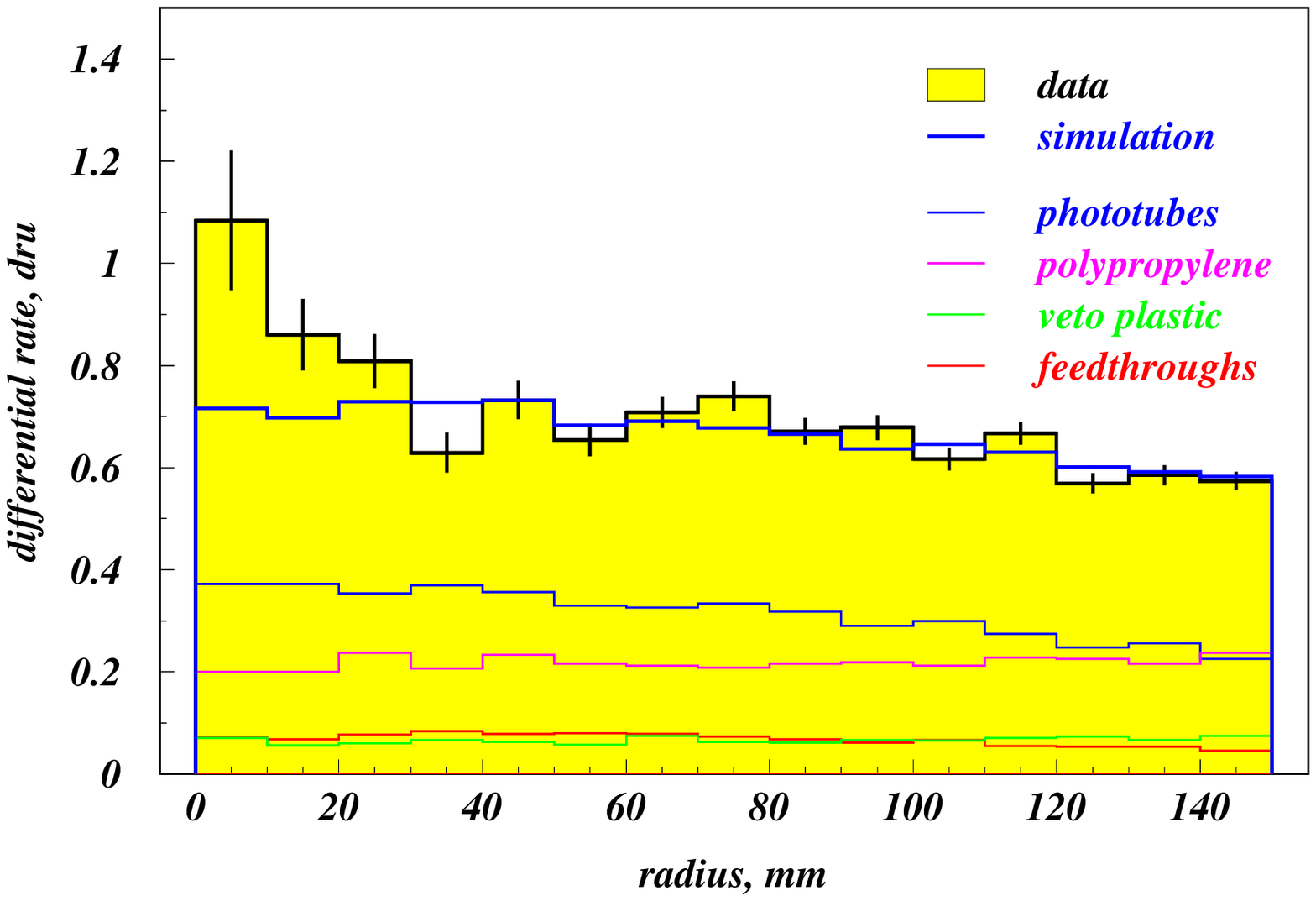}\\
    \includegraphics[width=7.8cm]{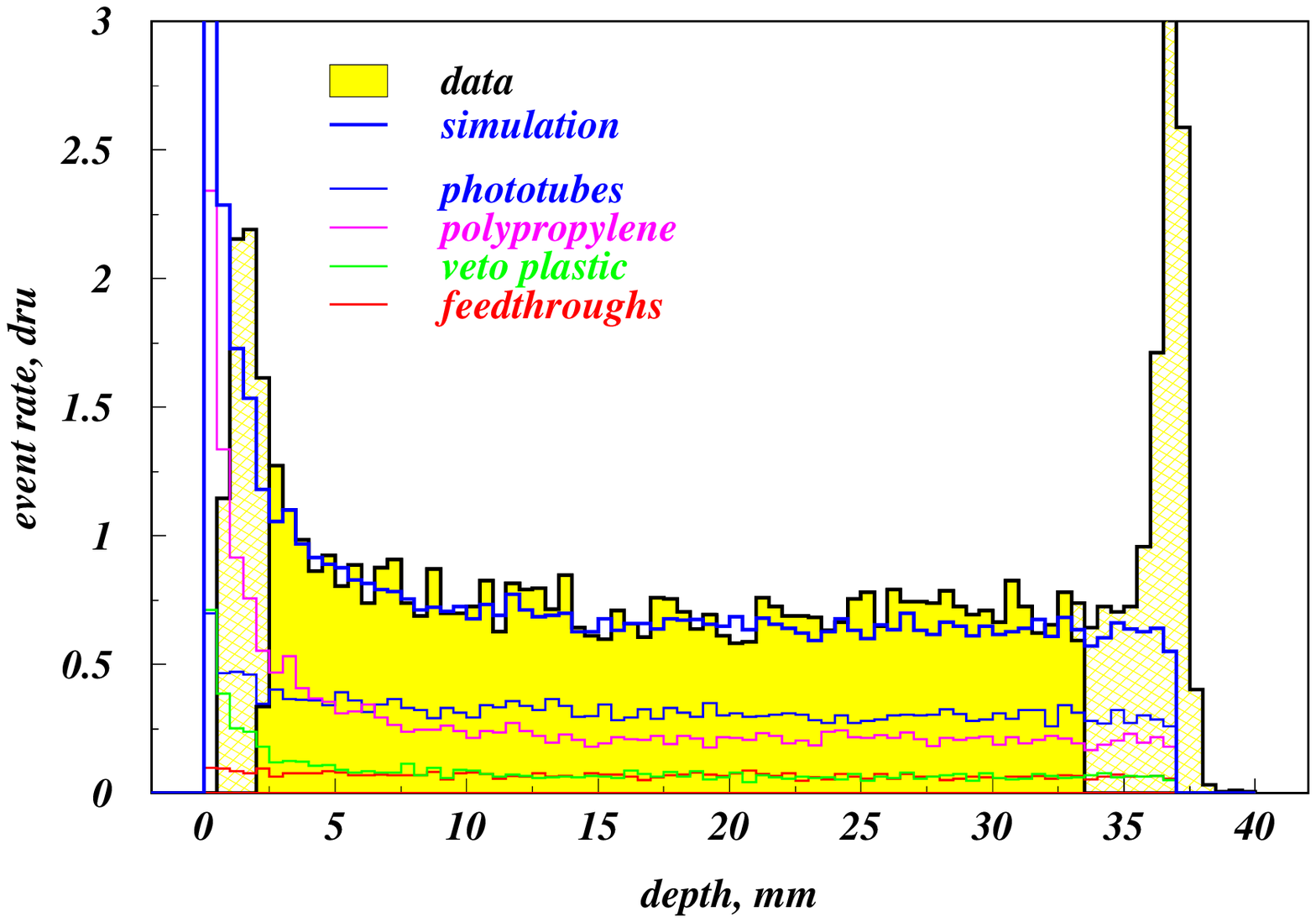}
  \end{center}
  \caption{\label{dru} Electron recoil background in ZEPLIN-III. Data
  are shown in yellow and simulated components are summed by the solid
  blue line, with individual contributions ordered by importance at
  low energy.  Top: energy spectrum in fiducial volume, showing good
  agreement below $\sim$150~keVee, the dynamic range of the dataset
  (the thin black histogram retains only unsaturated events); U/Th and
  Xe activation $\gamma$-rays are indicated. Middle: radial
  distribution below 100~keVee, confirming very good agreement beyond
  the central channel. Bottom: depth distribution for fiducial region
  (bright yellow) and full depth (hatched yellow); contamination of
  the cathode grid (37~mm) with radon progeny left over from the first
  run is clearly visible (but easily fiducialised away).}
\end{figure}

A 24-day long dataset was analysed to study the spatial and energy
distribution of electron recoil background. This followed a neutron
calibration by only a few weeks, and so isomeric transition (IT)
$\gamma$-rays from activation of the xenon are still visible (mainly
from $^{\rm 129m}$Xe, T$_{1/2}$=8.88~days, and $^{\rm 131m}$Xe,
T$_{1/2}$=11.8~days). The data were collected as part of the WIMP
search dataset and, for this analysis, the nuclear recoil signal
region remained blinded up to 40~keVee.\footnote{LXe energy deposits
are expressed in electron-equivalent `keVee' units when calibrated by
$^{57}$Co 122~keV $\gamma$-rays, or in `keVnr' when referring to
nuclear recoil energy.} However, this has little impact on the
electron recoil population. Events with single S1 and S2 pulses are
selected, which allows three-dimensional reconstruction of the
interaction point. A 6.5~kg fiducial region (150~mm radius, 32~mm
deep) is retained for analysis, with no significant quality cuts
applied. The energy variable used is a linear combination of S1 and S2
energy estimators which exploits the anti-correlation between these
two response channels (ionisation electrons may contribute either to
S2, if extracted from the track, or to S1, via recombination
luminescence). This allows for a better measurement of energy than
either channel independently. The dynamic range of the dataset is
limited, owing mainly to the 8-bit resolution of the
digitisers. Although a dual-range acquisition system is used, the
large S2 gain means that signal truncation becomes significant above
$\sim$150~keVee in the WIMP-search dataset.

The measured energy and spatial distributions are shown in
Figure~\ref{dru} -- along with a component-level breakdown from
simulated data calculated as described below. These contributions are
also listed in Table~I. The measured low-energy background is
0.75$\pm$0.05 events/kg/day/keVee (hereafter `dru', for `differential
rate unit'). Simulated and measured spectra are in good agreement up
to $\sim$150~keV and even beyond, despite the xenon activation which
is still prominent in this dataset. This represents a $\sim$20-fold
improvement over the first run. The simulation sum predicts
0.85$\pm$0.05~dru at low energy; as discussed further ahead, this is
based on measured contamination only, i.e.~excluding items for which
contamination levels could not be detected. Aside from this selection,
there is no scaling of the simulation results in that figure.

The spatial distributions are also in good agreement, with two
exceptions. The radial trend matches, within errors, over most of the
fiducial volume, but the rate is slightly higher than predicted within
the radius corresponding to the central photomultiplier (26.5~mm);
this is explained by higher than average contamination in this device
(but still within expected batch variability). The vertical
distribution is also reproduced closely, except for the spike observed
near 37~mm, the depth of the cathode wire grid. This is due to a
previous contamination with radon arising in the xenon purification
getters used prior to the first run. We subsequently found these to
emanate $\sim$1~atom/s, leading to significant implantation of
$^{210}$Pb in some detector surfaces. Notwithstanding, both energy and
spatial distributions of electron recoils are well reproduced by the
simulation within the fiducial volume.

\begin{table*}
\begin{center}
\caption{\small Component-level analysis of electron recoil background
and nuclear recoils from neutron elastic scattering in ZEPLIN-III. The
sum of simulated contributions (excluding components without measured
contamination, indicated in brackets) is labelled `SSR total'. The
additional rejection efficiency provided by the veto is indicated:
`ptag' for prompt tags, which remove $\gamma$-rays, and `dtag' for
delayed tags, which reject neutrons~\cite{ghag11} (weighted
averages are indicated under `SSR total'; these agree with the
measured tagging efficiencies). After veto tagging, the neutron event
expectation in a 1~year long dataset with typical signal acceptance is
some 10 times lower ($\simeq$0.4~events).}
  \begin{tabular}{l|c|cc|cc}
    \hline
    \hline
    Material & mass, kg & e-recoil, dru$^\dag$& ptag & n/year$^\ddag$ & dtag \\
    \hline
    Krypton-85           & --    & 0.007  & $\sim$0 & -- & -- \\
    Ceramic feedthroughs & 0.9   & 0.08   & 0.30  & 1.35 & 0.58  \\
    Photomultipliers     & 4.2   & 0.40   & 0.26  & 0.74 & 0.58  \\
    Rock (halite)        & --    &$\sim$0 & $\sim$0 & 0.53 & 0.58    \\
    Polypropylene shield & 1,266 & 0.25   & 0.04  & 0.10 & 0.58    \\
    Scintillator modules & 1,057 & 0.09   &$\sim$1& 0.03 & $\sim$1 \\
    Copperware           & $\sim$400 & ($<$0.1) & 0.10 & ($<$0.15) & 0.58 \\
    Lead castle          & $\sim$60,000 & 0.01 & 0.54 & $\sim$0 & 0.58 \\
    Radon-222            & 1~m$^3$& 0.03 & 0.19 & $\sim$0 & -- \\
    Muon-induced         & --    & --     &        & $\sim$0.3 & $\sim$1 \\
    \hline
    SSR total            &  & 0.86$\pm$0.05 & 0.28 & 3.05$\pm$0.5 & 0.58 \\
    SSR data             &  & 0.75$\pm$0.05 & 0.28 & n/a & -- \\
    \hline
    (FSR \cite{lebedenko09a} & & 14.5$\pm$0.5 & -- & (36$\pm$18)$^\ast$ & -- )\\
    \hline \hline
  \end{tabular}
  \end{center}
    {\small
    $^\dag$ events/kg/day/keVee at 10 keVee \\
    $^\ddag$ single scatters in 2,370~kg$\cdot$days over
    5--50~keVnr (unity detection efficiency)\\
    $^\ast$ FSR dataset expectation was 1.2$\pm$0.6 events in WIMP search box (net 128~kg$\cdot$days})
\end{table*}

\subsection{Plating of radon progeny on the cathode}

Although other primordial decay chains also include radon isotopes,
$^{222}$Rn ($^{238}$U chain) is the longest lived
(T$_{1/2}$=3.824~days). The issue of getter emanation had been
previously identified in ZEPLIN-II~\cite{alner07}, which relied on
continuous gas recirculation through the same getter model (SAES
PS11-MC500). Upon interruption of purification, the decay of the
$\alpha$ activity was measured at T$_{1/2}$=3.83$\pm$0.1~days,
supporting a $^{222}$Rn contamination~\cite{edwards09}. By then the
same getters had been used to purify ZEPLIN-III xenon before the first
run, which led to internal plating with $^{210}$Pb. This did not pose
a problem then, and we opted not to etch it away during the upgrade;
we replaced the getter with a new model (SAES PS4-MT3). The effect of
internal radon decay can be understood with reference to the chain
sequence:

\begin{footnotesize}
    \[ 
      {\rm \!\!\!\!\!\!\!\!\!\!\!
      ^{222}Rn \,(\alpha, 3.8d) \to 
      ^{218}Po \,(\alpha, 3.1m) \to 
      ^{214}Pb \,(\beta^-, 27m) \to
    }
    \]
    \[
      {\rm \!\!\!\!\!\!\!\!\!\!\!
	^{214}Bi \,(\beta^-, 20m) \to 
	^{214}Po \,(\alpha, 164\mu s) \to 
	{\bf ^{210}Pb \,(\beta^-, 22y)} \to
      }
      \]
      \begin{equation}
	{\rm \!\!\!\!\!\!\!\!\!\!\!
	  ^{210}Bi \,(\beta^-, 5d) \to 
	  ^{210}Po \,(\alpha, 138d) \to 
	  ^{206}Pb \,(stable)
	}
	\label{radon}
      \end{equation}
\end{footnotesize}

Once washed into the target the radon mixes uniformly in the LXe. The
first two $\alpha$ decays produce negatively charged atoms which
promptly lose excess electrons (the $\alpha$-particle drifts to the
cathode where it reduces to He). The next two $\beta^-$ decays create
$^{214}$Bi$^+$ and $^{214}$Po$^+$ ions which cannot transfer their
ionisation state to the rare gas and thus drift to the cathode.
Finally, the $\alpha$ decay of $^{214}$Po implants $^{210}$Pb
(T$_{1/2}$=22.3~yr) within the wire grid, which bottlenecks the chain.
This $^{210}$Pb activity is still present on the cathode in the second
run and has remained approximately stable at 19.4$\pm$0.3~decays/day
over several months (per each of $^{210}$Pb, $^{210}$Bi and
$^{210}$Po, as measured by the rate of $^{210}$Po nuclear recoils).

The electron-recoil background arising in the fiducial volume from
this contamination is insignificant in the context of the overall
radioactivity budget (and was not included in the simulation); the
same is true of nuclear recoils from the $(\alpha,n)$ reaction from
$^{210}$Po on the stainless steel wires.  In fact, these populations
can be useful as stable sources of ionisation; they allow, for
example, the ionisation yield for the deepest events to be monitored
as a function of time. The Bateman decay equations for the $^{210}$Pb
sub-chain indicate that the ratios of $^{210}$Pb, $^{210}$Bi and
$^{210}$Po become constant after $\sim$2~years (which is approximately
the period elapsed between the start of the two runs); when this
so-called transient equilibrium is reached, the relative spectrum of
decay products becomes constant, and so does the mean ionisation yield
per event. Monitoring the S2/S1 ratio for these low-energy events
allows, for example, the accuracy of the electron lifetime correction
to be checked.

\subsection{Photomultiplier $\gamma$-rays}

The original phototubes (2-inch ETEL D730Q) were purchased a decade
ago and had relatively high contamination levels (250~ppb U, 290~ppb
Th and 1350~ppm K), releasing 1,400~mBq per unit in $\gamma$-ray
activity. This dominated by far the background budget in the first run
(14.5~dru). For this reason a second run of the experiment with
upgraded devices was planned from the outset. A low-background product
(D766Q) was developed by ETEL in collaboration with the project to
enable pin-by-pin compatibility with the existing array. All materials
and sub-components were radio-assayed with HPGe and/or ICP-MS. The
best phototubes have 35~mBq in $\gamma$-rays, which was achieved
through a complete redesign of the device. Most of this activity is
concentrated at the rear of the envelope, away from the fiducial
volume. Simulations of background contribution to ZEPLIN-III take the
spatial distribution of activity into consideration. A value of
0.40~dru is predicted from internal $\gamma$-rays, which represents a
40-fold improvement over the previous array and makes this component
only marginally dominant. The equivalent simulations for the first run
reproduced data well, which gives added confidence in this
calculation. Backgrounds arising from the $\beta^-$ activity internal
to the phototubes, which are often neglected in this type of study,
are discussed in Section~\ref{kbetas}.

\subsection{Krypton-85 and other intrinsic backgrounds}

Several LXe-intrinsic backgrounds can affect this and similar
experiments. $^{210}$Pb from $^{238}$U and $^{222}$Rn chains and
$^{228}$Ra from the $^{232}$Th chain, for example, undergo low-energy
$\beta^-$ decay occasionally lacking significant x- and $\gamma$-rays
to help flag these events (`naked betas'). Tritium (T$_{1/2}$=12.3~yr,
$\beta_{max}$=19~keV) is produced cosmogenically and, on some
accounts, it could generate as much as $\sim$1~dru if the xenon is
stored on the surface for long periods \cite{mei09}. However, it is
reasonable to assume that these contaminants can be removed
effectively during xenon purification and, so long as radon emanation
into the detector can be controlled, they should not come to dominate
the background at low energy. We see no evidence for a substantial
presence of these in the second run of ZEPLIN-III.

More significant is the threat posed by $^{85}$Kr, long recognised as
a serious background for WIMP searches. $^{85}$Kr is an anthropogenic
$\beta^-$ emitter (T$_{1/2}$=10.76~yr, $\beta_{max}$=687~keV); its
presence in the atmosphere is mainly due to the reprocessing of spent
nuclear fuel. The pre-nuclear $^{85}$Kr/Kr ratio was as low as
3$\times$10$^{-18}$ in the early 1950s, but the present-day ratio is
$\sim$1$\times$10$^{-11}$~\cite{collon97}. 

The xenon used in ZEPLIN-III was sourced in the 1970s from an
underground origin. An accurate measurement of its $^{85}$Kr content
was obtained in 1997 from a 5.9~g sample \cite{gavrilyuk97}, from
which we derive an age-corrected $^{85}$Kr/Kr ratio of
1.5$\times$10$^{-12}$. The overall Kr content was subsequently reduced
by cryogenic distillation to $\sim$50~ppb Kr (by weight),
corresponding to 0.2~dru at low energy in the xenon target. This was,
however, a conservative estimate and the present data suggest much
lower contamination, which can be confirmed independently.

A minor $\beta^-$ decay (0.434\%, $\beta_{max}$=173~keV) is
accompanied by a 514~keV $\gamma$-ray from the $^{85}$Rb 9/2$^+$ level
(T$_{1/2}$=1.015~$\mu$s) \cite{sievers91}. This allows a
$\beta-\gamma$ delayed coincidence analysis, with the $\gamma$-ray
searched in the veto scintillator or in the LXe target. A 6-month
dataset revealed $\sim$9 such coincidences in the target, conforming
to the expected time delay distribution. This translates to
$4\!\times\!10^{-21}$ $^{85}$Kr/Xe ratio (in present-day xenon this
would represent a very low 150 ppt Kr). The decay rate is
0.007$\pm$0.002~dru, which is insignificant in ZEPLIN-III.

\subsection{Ceramics and copperware}

Along with the photomultipliers, two materials could make up a
significant fraction of the radioactivity budget of the experiment
from the outset: the $\sim$400~kg of copper (from which all structural
elements, vessels, fastening hardware parts, etc., were manufactured)
and ceramic feedthroughs. Other items were screened and their
locations chosen to minimise their impact: stainless steel parts
(e.g. vacuum ports), indium seals, LN$_2$ in the 30-litre reservoir.

For copper (OFHC C103), conservative upper limits of 0.5~ppb for U/Th
contamination were initially adopted, although lower values existed in
our database for similar material~\cite{ukdmc}. Simulation of this
contribution indicated an electron recoil background of 0.35~dru at
low energy. However, the measured energy and spatial distributions
show no evidence for such a significant rate, and we revised the upper
limit down to 0.1~dru for the primordial radioactivity.

Cosmogenic activation also produces several radio-isotopes, of which
$^{60}$Co (T$_{1/2}$=5.3~yr) will retain the highest activity a few
years after exposure to atmospheric muons ceases. We adopt a
production rate for $^{60}$Co in natural copper of $\sim$50
atoms/kg/day~\cite{mei09}. Since the exact exposure time on the
surface is uncertain we assume secular equilibrium, with a subsequent
cool-off period of 4 years (preceding the second run after deployment
underground). Simulation of the resulting $\gamma$-ray activity within
the copperware produces a negligible 0.01~dru in the fiducial volume.

There are nearly 100 ceramic feedthroughs for signal and high voltage
connections, weighing 900~g in total; these are distributed among the
xenon chamber baseplate and the vacuum vessel flange. They are alumina
based and their average contamination was measured at 105~ppb U,
270~ppb Th and 880~ppm~K. The $\gamma$-ray background is not severe
(0.08~dru), but they dominate the neutron background in the detector
as discussed later.

\subsection{Polypropylene shield and veto detector}

The ZEPLIN-III veto consists of a tight-fitting polypropylene shield
enveloping the main instrument, itself surrounded by 52 plastic
scintillator modules forming barrel and roof sections. The
gadolinium-loaded hydrocarbon moderates and captures internal neutrons
very efficiently; the Gd capture $\gamma$-rays are detected by the
scintillators. The neutron detection efficiency (delayed tagging in a
70~$\mu$s window) is 58\%; internal $\gamma$-rays are detected with an
average 28\% efficiency (prompt tagging in $\pm$0.2~$\mu$s
window). The prompt tag adds a further 2\% to the neutron
efficiency. A detailed discussion of the hardware and physics
performance is given elsewhere~\cite{akimov10b,ghag11}; there, we
analyse the backgrounds contributed to the WIMP target (a prime design
consideration) as well as background measurements in the veto detector
itself. A tonne of scintillator within the lead castle provides
confirmation of the local radiation environment and we find that the
veto measurements agree with the background measured within ZEPLIN-III
itself.

The readout (photomultipliers, electronics, cabling, etc.) are
external to the hydrocarbon shielding to mitigate neutrons, which is
achieved very successfully. These also have a negligible contribution
to the $\gamma$-ray background in the LXe target. However, owing to
their large mass, the polypropylene and the plastic scintillator do
contribute appreciably to the ZEPLIN-III budget.Contamination levels
have been measured for the polypropylene (1~ppb U, 1 ppb~Th and 5~ppm
K) and the plastic scintillator (0.2~ppb U, 0.1 ppb~Th and 0.2~ppm
K). These values are close to the sensitivity of the HPGe and ICP-MS
measurements of those samples, but this radiological contamination is
entirely consistent with the interaction rate recorded in the veto
itself. Furthermore, these have distinct energy and spatial
distributions in ZEPLIN-III, as shown in Figure~\ref{dru}, which
confirm these contributions: the energy spectra are generally softer
and they increase the event rate especially at the top of the xenon
target, in contrast with other dominant components, such as the
phototubes and ceramics, which are located below the xenon volume.

We note that radioactivity in the plastic scintillator does not
contribute to the overall background since the $Q$-value of the
radioactive decay, most of which is deposited locally, is nearly
always sufficient to trigger the veto (even if the highest energy
$\gamma$-rays are able to escape). On the other hand, the
polypropylene generates 0.25~dru in the xenon target and has a low
veto probability of 4\%.

\subsection{$\gamma$-rays inside lead castle}

The $\gamma$-ray flux inside the empty ZEPLIN-III castle includes
external contributions from the laboratory walls, natural as well as
cosmogenic activity in the shield, air-borne radon and its metallic
progeny plated out on the inner castle walls. $\gamma$-rays from
Boulby rock (halite) are attenuated by a factor of $\sim$10$^5$ by the
20-cm thick lead shield \cite{carson05} and are insignificant as a
background in the WIMP target. The activity of the $^{222}$Rn chain
measured with a commercial monitor inside the empty castle was
2.4~Bq/m$^3$, which is typical at Boulby. Simulation of $\gamma$-rays
from this activity leads to 0.03~dru in ZEPLIN-III. Further {\it in
situ} measurements with a HPGe detector place an upper limit on the
background from the castle (primordial and cosmogenic activity in the
bulk shield and inner walls). The castle walls and roof are made from
lead smelted into thin stainless steel containers to make 1.2-tonne
chevron-shaped blocks; the lead had been underground for some two
decades and is known to have low $^{210}$Pb content (it was used to
shield previous dark matter experiments at Boulby). The shield
flooring is OFHC copper, which was moved underground over a decade ago
as `new copper'. The HPGe spectra show only a vestigial amount of
$^{60}$Co. The combined contribution of the shield to ZEPLIN-III is
$\sim$0.01~dru.

\section{Rarer types of electron recoil event}
\label{odd}

\subsection{$\beta$- and $\gamma$-induced photomultiplier backgrounds}
\label{kbetas}

A significant amount of potassium is involved in sensitising internal
photomultiplier surfaces, namely the bialkali photocathodes. The
$^{40}$K thus located close to the fiducial volume can generate a
response in the detector in several ways, with the 1,461~keV EC
$\gamma$-rays being perhaps the most obvious. However, the 89\%
branching ratio $\beta^-$ decay ($\beta_{max}$=1,311~keV) can be as
problematic (this represents 30~mBq in the D766Q). VUV-sensitive
phototubes have thin quartz windows ($\sim$1~mm), allowing some
internal $\beta$-particles to reach the LXe. Others will generate
bremsstrahlung photons in the quartz. Cherenkov photons are also
produced in the quartz and can be detected in coincidence with the
prompt scintillation. Other luminescence processes arise internally in
the phototubes, such as `dynode glow' from electron impact, which is
signal coincident. Besides contributing some 10--20\% to the detector
trigger rate, these $\beta$-induced events can lead to backgrounds
which are harder to discriminate than single scatters in the fiducial
volume. These might combine, for example, bremsstrahlung photons
interacting in the main volume with Cherenkov light from the quartz,
thus decreasing the observed S2/S1 ratio of the fiducial interaction.
Cherenkov emission from Compton electrons created by $\gamma$-ray
backgrounds may be even more damaging as fewer photoelectrons may be
involved in this instance.

\begin{figure}[ht]
  \begin{center}
  \includegraphics[width=7.8cm]{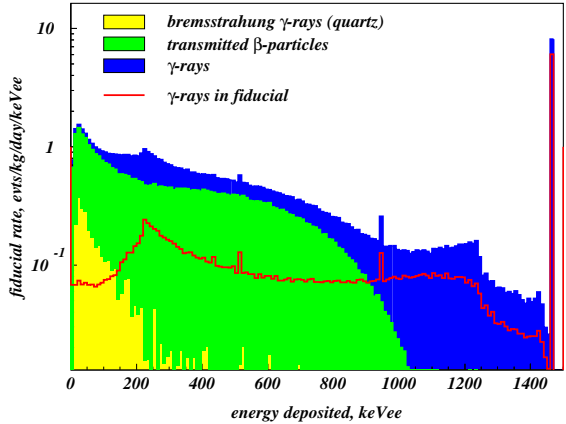}
  \end{center}
  \caption{Energy spectrum in LXe from simulated $^{40}$K decays
    internal to central photomultiplier; curves are scaled
    ($\times$31) for correct fiducial rate from this contribution. The
    solid histograms are cumulative and represent interactions above
    the phototube window; these include bremsstrahlung photons
    generated in the quartz, transmitted $\beta$-particles and
    non-bremsstrahlung $\gamma$-rays. The spectrum above the cathode
    is shown in red, mostly due to 1,461~keV $\gamma$-rays.}
  \label{kgen}
\end{figure}

Figure~\ref{kgen} illustrates some of these processes; it shows Monte
Carlo data for $^{40}$K decays generated internally in the central
phototube in ZEPLIN-III. The quartz windows are curved and $<$1~mm
thick. The blue histogram (cumulative) represents energy deposits
everywhere in LXe whereas the red histogram refers to the active
region above the cathode only (mainly $\gamma$-rays). Transmitted
$\beta$-particles and bremsstrahlung in the quartz window clearly
dominate at low energies for interactions just above the phototube.
In the fiducial volume, some 10~mm away, these contributions become
negligible.

\begin{figure}[ht]
  \begin{center}
  \includegraphics[width=7.8cm]{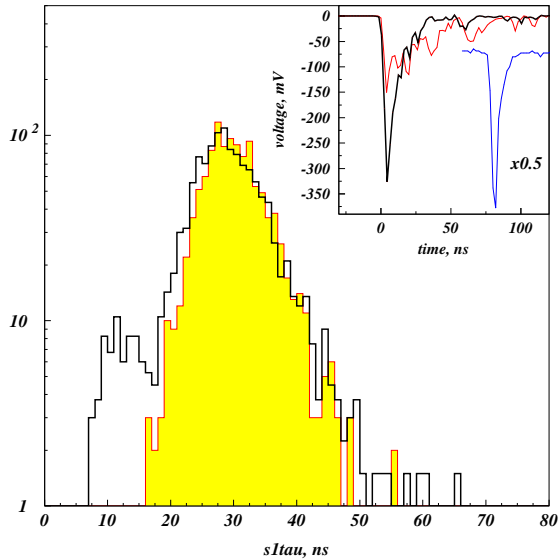}
  \end{center}
  \caption{Distribution of photoelectron mean arrival times ($\tau$)
  in S1-like pulses for fiducial interactions (yellow histogram) and
  for S1-only events occurring between the photomultiplier windows and
  the cathode grid. The apparent energies are 30--60~keVee for both
  histograms. A population of fast pulses is clearly visible, which we
  attribute to Cherenkov photons generated in the phototube windows by
  internal $\beta^-$ emission. Inset: array time response for three
  events with 30~keVee reconstructed energy (approximately
  55~photoelectrons); the red trace is a fiducial event with
  $\tau$=35~ns; the black trace, probably a scintillation event
  contaminated by significant Cherenkov light, is chosen from the
  black histogram with $\tau$=13~ns. The pulse in blue has $\tau$=7~ns
  and occurs in only one channel (the other cases both trigger 7
  channels).}
  \label{cherenkov}
\end{figure}

As mentioned above, Cherenkov light generated inside the quartz may
also contribute to S1, yielding potentially tens of photoelectrons.
For example, a 1~MeV electron -- e.g.~a $^{40}$K $\beta$-particle or a
Compton electron from the 1,461~MeV $\gamma$-ray -- produces $\sim$100
photons/mm of quartz within the spectral response of these
photomultipliers; some generate photoelectrons even without leaving
the window, while others suffer total internal reflection at the
liquid surface and strike the array again; therefore, the
photoelectron yield can be significantly higher than suggested by the
nominal QE. This optical signal can both increase the size of a
time-coincident scintillation pulse as well as reduce its apparent
time constant, threatening single-phase (scintillation only) as well
as two-phase detectors. We see evidence of Cherenkov emission in
ZEPLIN-III. This is illustrated in Figure~\ref{cherenkov}, where the
pulse timing parameter $\tau$, which measures the mean arrival time of
the S1 signal, is shown for fiducial scintillation pulses and for
sub-cathode events (no S2) with the same range of apparent energies
(30--60 keVee). The latter are mainly due to $\beta$-particles
transmitted through the phototube windows -- but a faster population
is also clearly seen which we attribute to Cherenkov emission. We can
rule out an explanation based on nuclear recoils on the underside of
the cathode grid, which would indeed be faster than electron recoils
and yield no ionisation, by comparing with the corresponding timing
spectrum for the upper grid surface. We conclude that pulses in this
population are indeed faster than nuclear recoil scintillation. More
extreme examples can be identified if signals triggering in a single
channel are included (such as the blue pulse shown in inset), but in
this instance it is harder to prove an optical, rather than
electrical, origin.

Such Cherenkov processes can be dangerous if associated with
low-energy electron recoils in the fiducial volume. Besides Compton
electrons generated by background $\gamma$-rays, this could involve,
more generally, $\beta^-$ decays with coincident $\gamma$-rays (many
examples exist in the U and Th chains). In ZEPLIN-III, the timing
distribution of fiducial events with higher than average scintillation
(lower S2/S1 ratio) at energies above 30~keVee does not correlate with
shorter than average pulses; rather, it is fully compatible with that
of electron recoils -- although this conclusion stems from a
relatively small exposure. In any case, this effect should be
carefully assessed in future noble liquid instruments where
scintillation is detected by photomultipliers.

For this and other reasons optical cross-talk between phototubes must
be avoided, so that these effects remain localised to a single channel
and are therefore easier to identify. It is also important to reduce
the amount of LXe allowed around their envelopes to minimise
scintillation from internal $\beta$-particles, and to prevent those
photons from reaching the photocathode either by external or internal
paths. The copper structures which house the ZEPLIN-III phototubes
allow a thicker LXe layer around their envelopes than with the
original ones, since the upgrade photomultipliers are slightly
narrower. However, there is no significant direct path into the main
volume and the quartz envelopes were metallised to prevent light from
reaching the photocathode directly.

\subsection{Multiple $\gamma$-ray interactions: MSSI events}
\label{mssi}

Multiple interactions where one vertex occurs in a detector region
which is optically coupled to the phototube array but from which no
charge can be extracted can lead to a challenging background.
Multiple-scintillation single-ionisation (MSSI) events produce lower
S2/S1 ratios than single fiducial scatters since the multiple S1
pulses are time-coincident but there is only one S2. This was
identified early on as a background mechanism in two-phase
detectors. In ZEPLIN-III an initial assessment during construction
showed that efficient rejection based on S1 light pattern and S1-S2
vertex consistency would be necessary~\cite{araujo06}. In the first
run this event topology became perhaps the most challenging, in
particular because these populations cannot be reproduced accurately
with calibration point sources.

A detailed calculation is difficult when the light collection for the
dead vertex is not known precisely. In ZEPLIN-III, this is the case in
two regions which yield no charge: i) the reverse field region below
the cathode grid, where light collection is high but variable due to
the proximity to the photomultiplier windows; ii) the peripheral
region near the side walls, where the light collection is low and also
uncertain. In any case, it is instructive to assess what effect these
events will have on discrimination if not successfully
rejected. Consider two $\gamma$-ray interactions, one in the fiducial
region and one in a dead region, which we assume to have equal light
collection and a constant spectrum of energy deposited (as suggested
by the Klein-Nishina formula for low-angle scattering of high energy
$\gamma$-rays). The energy is reconstructed from the coincident S1
components, but there is only one S2. The number of such double
scatters adding up to energy $E$ is $P(E)dE = P_0^2 E\,dE$, where
$P_0$ is the single-scatter probability per unit energy. So, this
effect becomes more frequent with increasing reconstructed energy, as
might be expected. The discrimination parameter, S2/S1, will suffer
most when the `good' vertex produces the smallest detectable S2 signal
($E_{th}$) and S1 takes its apparent energy from the dead vertex only
(S2/S1$>$$E_{th}/E$). However, the ($x,y$) positions recovered from S1
and S2 independently are likely to be inconsistent for these events,
leading to their probable rejection. Therefore, the good S1 vertex
must produce a minimum viable signature and, in practice,
$E_{th}\!\sim\!2.5$~keVee, the S1 energy threshold. This trend is
indeed observed in our data before these events are removed by the
vertex reconstruction algorithm. A similar calculation can be done for
neutron interactions. Assuming, very approximately, an exponential
recoil spectrum with characteristic energy $E_0$, then the spectrum of
double-vertex events is $P(E)dE\!\propto\!E\exp({-E/E_0})\,dE$
(i.e.~the relative number of such events still increases with $E$).

\section{Radioactivity neutrons}
\label{neutrons}

The agreement between data and simulations obtained for the spectrum
of electron recoils at low energy confirms the radiological
contaminations adopted for each component; these were subsequently
used to derive neutron production rates in those
materials. Muon-induced showers, in particular photo-production in
electromagnetic cascades, are the only noteworthy source of
non-radioactivity neutrons. This contribution is not dominant in a
detector of the size of ZEPLIN-III~\cite{araujo08,lindote09} and, in
addition, these neutrons are vetoed effectively. The dominant sources
of fast neutrons are, therefore, spontaneous fission (mostly of
$^{238}$U) and ($\alpha$,n) reactions arising from the U and Th
chains.\footnote{An $\alpha$ activity of 10~Bq is present in the
polypropylene due to $^{152}$Gd, but this is not a significant source
of neutrons ($E_\alpha$=2.8~MeV).}

Absolute neutron spectra were calculated with SOURCES-4A and
-4C~\cite{wilson99,wilson02} with ($\alpha$,n) cross-sections extended
to 10~MeV as described in Ref.~\cite{carson04} and complemented with
EMPIRE2.19~\cite{herman07} cross-sections and branching ratios for
transitions to excited states~\cite{tomasello08}. The ($\alpha$,n)
reaction dominates in light materials, producing, e.g., nearly all
neutrons emitted by the Boulby rock (considered pure NaCl here), very
few in copper ($<$10\%) and practically none in lead. Yields in pure
elemental materials are generally in good agreement with experimental
data, but it should be noted that the exact composition, uniformity
and granularity of mixtures in detector components is a cause of
uncertainty due to the very short range of $\alpha$-particles.

Primary neutron spectra are generated isotropically from an
appropriate position in the GEANT4 simulation geometry and tracked to
the LXe target. In the case of the photomultipliers this is done at
sub-component level, since some materials are further away from the
fiducial volume than others. Each photomultiplier emits just over 2
neutrons per year; this is dominated by a small amount of borosilicate
glass still present in this model; although there is scope for further
improvement, this value represents a 50-fold reduction relative to
those used in the first phase of the experiment.

The most challenging calculation is the contribution from the
laboratory rock. It is computationally intensive and requires accurate
modelling of the shield: in spite of the high overall attenuation
factor of $\sim$10$^5$, any gaps for pipework, cabling and alignment
tolerances need to be taken into account. To ensure an accurate result
a detailed CAD solid model of the veto and shield was imported into
the GEANT4 model. The Boulby rock has 65~ppb U and 130~ppb Th assumed
in secular equilibrium \cite{smith05}. Neutrons are generated in the
rock to a depth of 3~m and tallied initially on a test surface in the
laboratory. The simulated integral flux is adjusted at this stage to
measurements at Boulby \cite{tziaferi07}, including re-entrant
(i.e.~backscattered) neutrons. The differential spectrum is released
with a cosine bias from a generator sphere containing the experiment
to generate an isotropic and uniform flux within it.

\begin{figure}[ht]
  \begin{center}
  \includegraphics[width=7.8cm]{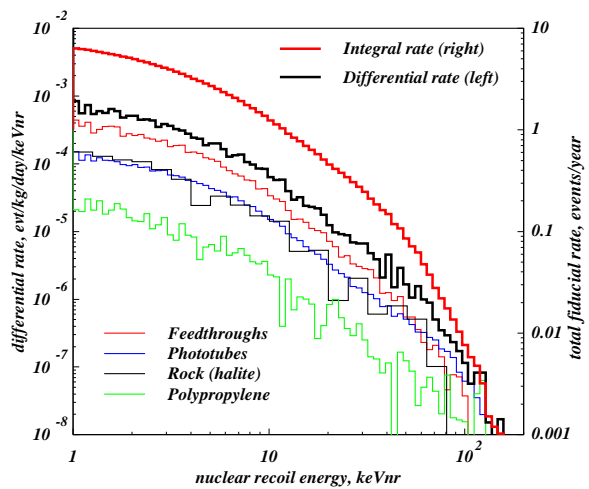}
  \end{center}
  \caption{\label{ndru} Simulated differential and integral rates of
    neutron single elastic scattering in the fiducial volume of
    ZEPLIN-III.}
\end{figure}

Yearly rates of single scatters in 5--50~keVnr in the fiducial volume
and the component-level breakdown are given in Table~I and shown in
Figure~\ref{ndru}. The total of 3.05 events per year is dominated by
the ceramic feedthroughs (44\%), followed by the photomultipliers and
rock neutrons. We note that the background expectation predicted for a
one-year dataset with realistic signal acceptance is considerably
lower; typical signal detection efficiencies (including 90\%
operational duty cycle, 50\% acceptance in S2/S1 discrimination
parameter, a further 25\% loss from software selection cuts), bring
this down to 1.0 n/year; the rate in anti-coincidence with the veto is
60\% lower, at~0.4~n/year. Finally, the detection threshold is likely
to be a little higher than 5~keVnr, so a value just below 0.3~n/year
is expected. This is very close to the original goal of the
experiment. The veto-coincident data analysed so far contains zero
delayed coincidences (dtag) within the signal acceptance region, but
this constrains the latter event rate only to $<$3~n/year at 90\% CL
(note that vetoed events have not been subject to a blind analysis
since they contains no information on the presence of a signal).

The event rates discussed above are for single scatters in the 6.5~kg
fiducial volume with no additional interactions (elastic or otherwise)
in the 12.5~kg active volume. Where more than one elastic scatter
occurs, the low S2/S1 ratios (and short scintillation time constant)
characteristic of nuclear recoils are essentially preserved. Although
the event can still be rejected from the multiple S2 pulses, it is
often useful to establish the distribution of scattering
multiplicity. This may provide additional constraints to the expected
rate of single scatters, which are an irreducible background for WIMP
searches. In ZEPLIN-III, some 50\% of events with energies
reconstructed in the 5--50~keVnr region are single interactions. The
mean scattering length varies quite significantly with neutron energy,
but it is much less sensitive as a function of recoil energy
($\lambda\!\sim\!30$~cm). For systems much larger than this value,
self-shielding can render the fraction of single scatters very
small. Multiple scatters can be identified by additional S2 pulses in
two-phase systems (although they will become MSSI events if the second
vertex yields no charge). However, the time-of-flight for
radioactivity neutrons should not be neglected as a valuable
discriminant. The mean time between scatters (above threshold) is only
$\sim$15~ns in ZEPLIN-III, but this effect can be readily seen in the
structure of S1 pulses during neutron calibration, which points to its
usefulness in large single-phase detectors.

\section{Discussion}
\label{conclusion}

In this study we analysed electron and nuclear recoil backgrounds of
radiological origin in the ZEPLIN-III experiment now operating in its
second phase at Boulby. Energy spectra and spatial distributions are
presented for electron recoils in the fiducial volume. The low energy
rate is 0.75$\pm$0.05~evt/kg/day/keVee, which represents a 20-fold
improvement relative to pre-upgrade levels. A further 28\% reduction
is possible due to the new veto detector installed around the WIMP
target. Significantly, these vetoed events can be studied without
compromising a blind analysis of the signal region.

The electronic background achieved in phase~II meets the original goal
of the ZEPLIN-III programme. It implies a leakage of 2--3 events/year
for a discrimination power of 1:7,800 (the value achieved in the first
science run). Whether this can be matched in the second run depends on
the optical performance of the photomultipliers, which unfortunately
is not ideal.

A long campaign of component-level radio-assays (and Monte Carlo
simulations to determine their contribution to the radioactivity
budget) led to good matches to the measured distributions. We
confirmed that the photomultipliers and the polypropylene shield
dominate, while copper and $^{85}$Kr, identified early on as possible
concerns, contribute very little. Similar studies for other
experiments, namely XENON100~\cite{aprile11}, have been equally
successful, confirming that this methodology produces accurate
results.

The good agreement achieved for electron recoils also confirms the
primordial U/Th contamination of key components, lending more
certainty to the calculation of neutron backgrounds from natural
radioactivity. The total rate of single elastic scatters in the
fiducial volume is 3~events/year in the energy range 5--50~keVnr,
which translates to 0.4~events/year in anti-coincidence with the veto
for a realistic signal acceptance. This will not limit the sensitivity
of the experiment. We also pointed out that neutron time-of-flight can
be important to establish scatter multiplicity in large single-phase
experiments.

We examined event topologies which challenge the high discrimination
power which can be achieved for single-vertex electron
recoils. Cherenkov light produced in photomultiplier windows by
internal $\beta^-$ activity or by Compton electrons from background
$\gamma$-rays can be detected in coincidence with $\gamma$-ray
fiducial scatters, thereby making scintillation pulses appear larger
as well as faster -- i.e.~more similar to nuclear recoils. Dynode glow
is another cause of signal-coincident light, although it should affect
both S1 and S2 pulses. In this instance photon emission is delayed
with respect to the original optical stimulus depending on the input
optics configuration and its voltage bias. These mechanisms were
identified several decades ago (see, e.g.,~\cite{wright83} and
references therein). In the context of dark matter searches, tests
conducted at Imperial College in the early 1990s confirmed that dynode
glow and Cherenkov light were observed when two photomultipliers face
each other without an intervening scintillator~\cite{li91};
subsequently, the NAIAD experiment at Boulby documented a population
of fast noise events ($\tau\!<\!100$~ns) which were attributed to
dynode glow (see Fig.~1 in Ref.\cite{alner05}). More recently, a WIMP
search with CaF$_2$(Eu) scintillator at Kamioka Observatory found that
the sensitivity below 10~keVee was limited by Cherenkov photons
produced in light guides by Compton electrons caused by background
$\gamma$-rays~\cite{shimizu06}. This type of event may be difficult to
identify if the light pattern is not too peaked in a single
channel. We note that these effects should also concern single-phase
experiments such as XMASS~\cite{suzuki00}, MiniCLEAN
~\cite{mckinsey07} and DEAP-3600~\cite{boulay08}.

We discussed also $\gamma$-ray MSSI events and how these degrade the
discrimination power of two-phase detectors. They pose a real
challenge to current experiments and their mitigation should attract
significant design effort in future systems. Whereas single-vertex
electron recoils have a steep distribution in the S2/S1 discrimination
parameter, and their leaking into the nuclear recoil region is
progressive, this is not the case for MSSI events. Fortunately, both
the frequency and severity of these events decreases for lower
energies, where more signal is expected.

Under the assumption that the discrimination power eventually observed
is not hindered significantly by poor photomultiplier performance or
incomplete rejection of MSSI events, ZEPLIN-III should achieve a
sensitivity of $\sim$1$\times\!10^{-8}$~pb$\cdot$year to the scalar
WIMP-nucleon elastic cross-section.

\section{Acknowledgements} 

The UK groups acknowledge the support of the Science \& Technology
Facilities Council (STFC) for the ZEPLIN-III project and for
maintenance and operation of the Boulby underground laboratory.
LIP-Coimbra acknowledges financial support from Funda\c{c}\~{a}o para
a Ci\^encia e a Tecnologia (FCT) through project grant
CERN/FP/116374/2010 and postdoctoral grants SFRH/BPD/27054/2006,
SFRH/BPD/47320/2008 and SFRH/BPD/63096/\-2009. The ITEP group
acknowledges support from the Russian Foundation of Basic Research
(grant 08-02-91851 KO\_a) and Rosatom (contract H.4e.45.90.10.1053
from 03-02-2010). We are also grateful for support provided jointly to
ITEP and Imperial from the UK Royal Society.  ZEPLIN-III is hosted by
Cleveland Potash Ltd (CPL) at the Boulby Mine and we thank CPL
management and staff for their long-standing support. We also express
our gratitude to the Boulby facility staff for their dedication. The
University of Edinburgh is a charitable body registered in Scotland
(SC005336).

\bibliographystyle{elsarticle-num}
\bibliography{HAraujo.bib}

\end{document}